\documentclass[10pt,pre,aps,twocolumn,showpacs,superscriptaddress,floatfix]{revtex4}
\usepackage{graphicx}

\newcommand{\be}{\begin{equation}}
\newcommand{\ee}{\end{equation}}
\newcommand{\bea}{\begin{eqnarray}}
\newcommand{\eea}{\end{eqnarray}}

\begin{document}

\title{Directed-loop Monte Carlo simulations of vertex models}

\author{Olav F.~Sylju{\aa}sen}
\affiliation{NORDITA, Blegdamsvej 17, Copenhagen {\O}, DK-2100
Denmark} \email{sylju@nordita.dk}
\author{M. B.~Zvonarev}
\affiliation{\O rsted Laboratory, Niels Bohr Institute for APG,
Universitetsparken 5, Copenhagen {\O}, DK 2100, Denmark}
\email{zvonarev@fys.ku.dk}

\date{\today}

\preprint{NORDITA-2004-004}

\begin{abstract}
We show how the directed-loop Monte Carlo algorithm can be applied
to study vertex models. The algorithm is employed to calculate the
arrow polarization in the six-vertex model with the domain wall
boundary conditions (DWBC). The model exhibits spatially separated
ordered and ``disordered'' regions. We show how the boundary
between these regions depends on parameters of the model. We give
some predictions on the behavior of the polarization in the
thermodynamic limit and discuss the relation to the Arctic Circle theorem.

\end{abstract}

\maketitle

\section{Introduction}
Vertex models have a long and distinguished history in physics.
Their fame is intimately connected to the concept of
integrability, and the exact solutions of the
six-vertex~\cite{Lieb} and the eight-vertex~\cite{Baxter-82}
models with periodic boundary conditions (PBC) are indeed
milestones in physics. Despite being exactly solvable, there are
questions about these models that cannot easily be answered. An
example is the influence of boundary conditions on correlation
functions. While boundary conditions are not normally important in
the thermodynamic limit, they have a profound influence on the
vertex models. Exact studies, made for the six-vertex model with
the domain wall boundary conditions (DWBC)~\cite{KBI-93} show this
in particular. These studies were restricted to certain points in
the phase diagram, and involve rather sophisticated mathematical
methods. It is thus appropriate to complement them with Monte
Carlo simulations.

The purpose of this article is to demonstrate that the
directed-loop Monte Carlo algorithm developed for quantum spin
systems~\cite{SS} can be used as an effective tool to study vertex
models. The discussion of the algorithm will be kept general, but
when demonstrating its use we will focus on the six-vertex model
with the DWBC, a model which is difficult to simulate using other
known Monte Carlo algorithms.

\section{Monte Carlo algorithm}
In a vertex model, each vertex have edges with an Ising-like
variable, an arrow, that points either away from or into the
vertex. The arrangement of arrows around the vertex determines the
vertex weight. Two vertices are joined by their common edge,
sharing the arrow on the edge. In general there are no
restrictions on which vertices are joined, however for traditional
vertex models nearest-neighbor vertices are joined together. The
Monte Carlo algorithm discussed here always flips two (or zero)
arrows on a vertex, thus it is limited to models where an even
number of arrows are pointing away from each vertex. Most vertex
models of interest obey this rule.

\begin{figure}
\includegraphics[clip,width=6cm]{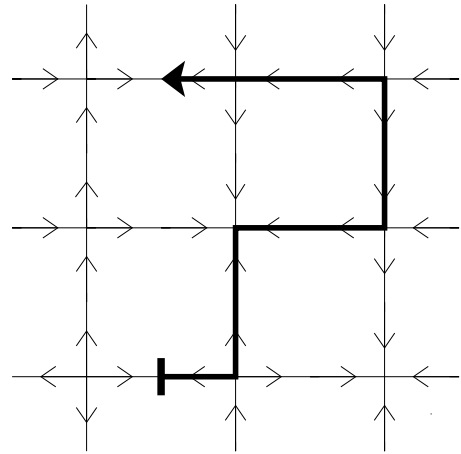}
\caption{Illustration of the directed-loop algorithm. Vertex edges
are drawn with two arrows allowing the discontinuity at the head
and tail of the loop to be shown. The thick line shows the loop path
along which the arrows has been flipped. The loop closes when the
loop head (thick arrow) hits the loop tail (vertical
bar).\label{algorithm}}
\end{figure}

In visualizing the directed-loop Monte Carlo algorithm, originally
developed for quantum
systems~\cite{SS}, it is helpful to cut every edge into two pieces,
each piece having an arrow belonging to a specific vertex,
Fig.~\ref{algorithm}.
\begin{figure}
\begin{center}
\includegraphics[clip,width=8.5cm]{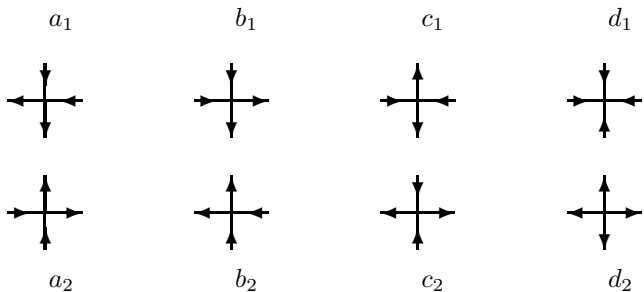}
\end{center}
\caption{The vertices of the eight-vertex model and their
statistical weights.} \label{vertices}
\end{figure}
For a valid vertex configuration the arrows on the two parts of an edge
must have the same orientation. The directed-loop
algorithm is as follows: Pick a random vertex $v_1$ and a random
edge belonging to that vertex. Based on these choices select in a
probabilistic manner another edge belonging to $v_1$ and name
that the out-edge. Then flip the arrows on both the part of the
in-edge and the part of the out-edge belonging to $v_1$. This
introduces two discontinuities in the arrow configurations on the
edges, one on the starting in-edge and another one on the out-edge.
The new configuration is thus not an allowed vertex configuration.
To repair this, the out-edge discontinuity is moved by repeating
the procedure on the vertex connected to the out-edge $v_2$, this
time using the out-edge of $v_1$ and the in-edge on $v_2$. The
process is stopped when the out-edge selected is the starting
edge, thus healing all discontinuities. In this way arrows are
flipped as a loop is constructed, and a new allowed vertex configuration is arrived at when the loop closes.

In order to determine the probabilities for selecting out-edges and to see how detailed balance is satisfied one needs to consider also the probability for the reverse update. The reverse update consists of traversing the same loop in the opposite direction while flipping arrows back.
As is explained in detail in Ref.~\cite{SS}, detailed
balance is satisfied for the whole loop construction, if detailed
balance is satisfied in each edge selecting step, for which the
criterion is as follows: Let $w$ be the weight of the vertex $v$
before edge-flips, then the probability $P(v,i \to o)$ for exiting
at the out-edge $o$, given that the in-edge is $i$, should satisfy
\be
\label{detbal1}
     w P(v, i \to o) = w^\prime P(v^\prime, o \to i),
\ee where $w^\prime$ is the weight of the vertex $v^\prime$
obtained by flipping the arrows on edges $i$ and $o$ belonging to
the vertex $v$.
Notice that $P(v^\prime, o \to i)$, on the right hand side, describes an edge-selecting step in the reverse update process
where the loop is traversed in the opposite direction to that described
on the left hand side.
The loop construction should not terminate in the edge-selecting step, thus
\be
\label{detbal2}
     \sum_o P(v,i \to o) = 1,
\ee
where the sum is taken over all possible out-edges, including
the in-edge $i$.

\begin{figure}
\begin{center}
\includegraphics[clip,width=8cm]{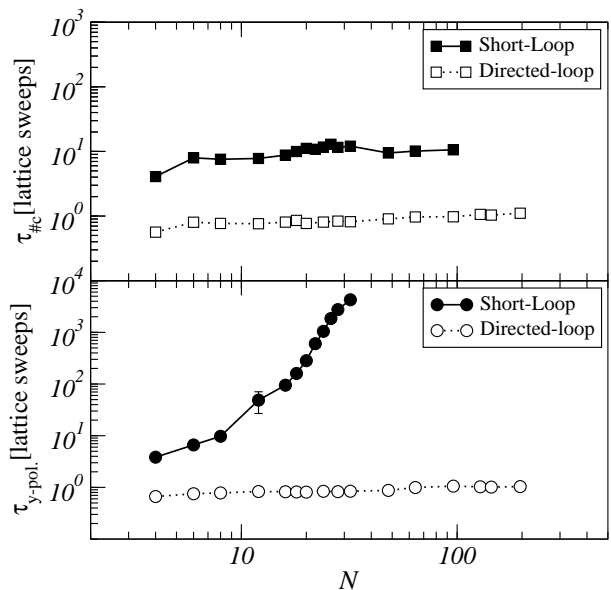}
\end{center}
\caption{ Integrated autocorrelation times for number of $c$-type
vertices (upper panel) and the total polarization in the
$y$-direction (lower panel) for the directed-loop algorithm (open
symbols) and the short-loop Barkema-Newman algorithm (closed
symbols). The data shown is for the symmetric six-vertex model on
an $N \times N$ square lattice with PBC and vertex weights $a=b=2$
and $c=1$.  \label{Fig:Autocorr}}
\end{figure}

This algorithm resembles closely the ice model algorithm invented by Rahman and Stillinger~\cite{Rahman}, generalized to arbitrary couplings by Barkema and Newman~\cite{BN}.
In fact, at the point in parameter space where all vertex weights are equal our algorithm is identical to the long-loop version of the ice model algorithm. However away from this point,
Barkema and Newman's algorithm involves accepting or rejecting
the loop after it has been constructed. The directed-loop algorithm has
no such accept/reject step. A comparison of integrated autocorrelation
times for the directed-loop algorithm and the short-loop algorithm of Barkema and Newman are shown in Fig.~\ref{Fig:Autocorr}.
The autocorrelation times are measured in units of lattice sweeps. One lattice sweep corresponds to a number of elementary loop moves such that on average each vertex on the lattice have been visited twice. In
defining visited we include parts of the loop where the loop bounces off a
vertex (relevant for the directed-loop algorithm) and
the neck part of short-loops. Neither the bounces nor the short-loop-necks
contribute to changes in the vertex configuration.
However they are intrinsic parts of the algorithms and
requires computer effort, and should therefore be accounted for.

The upper panel of Fig.~\ref{Fig:Autocorr} shows integrated autocorrelation times of the observable counting the number of $c$-type vertices in each configuration. This observables was chosen to compare with the performance results in Ref.~\cite{BN}. While the integrated autocorrelation times are larger for the short-loop algorithm the scaling with system size appears to be equal for both
algorithms. The lower panel shows integrated autocorrelation times for the total arrow-polarization in the y-direction. These scales much worse for the short-loop algorithm than for the directed-loop algorithm. This is to be expected from the fact that most loops accepted in the short-loop algorithm are small, while large loops that wind around the boundary of the lattice is needed to change the total polarization. These are not suppressed in the directed-loop algorithm, thus leading to better performance.

The Eqs.~(\ref{detbal1}) and (\ref{detbal2}) form several coupled sets which
in most cases are under-determined. There are thus many solutions
for the out-edge selection probabilities $P$.
Some general solutions and analysis of their efficiency for
different quantum systems were reported in Ref.~\cite{Olav}.
Here we employ the solution B in Ref.~\cite{Olav} to the
eight-vertex model, but solutions for higher-vertex models are not
hard to find as well. The allowed vertices for the eight-vertex
model and their statistical weights are shown in
Fig.~\ref{vertices}. To shorten notation, we consider the
so-called symmetric case: the statistical weights, $a,$ $b,$ $c,$
and $d,$ of the allowed states are assumed to be invariant under
the simultaneous reversal of all arrows. The generalization of the
algorithm to the non-symmetric case can be performed easily.

Let $W_1, \dots ,W_4$ be the vertex weights $a,b,c,d$ of the
eight-vertex model ordered so that $W_1 \geq W_2 \geq W_3 \geq
W_4$. Then the probability for picking the out-edge on a vertex
with weight $W_i$ resulting in a new vertex weight $W_j$ after
flipping arrows is $t_{ij}/W_i,$ where $t_{ij}=t_{ji}$
and the non-zero entries of the $4\times4$ matrix $t$ are
\bea
    t_{12} & = & (W_1 + W_2 -W_3 -W_4)/2 , \nonumber \\
    t_{13} & = & (W_1 - W_2 +W_3 -W_4)/2 , \label{rules} \\
    t_{23} & = & (-W_1 + W_2 +W_3 +W_4)/2 , \nonumber \\
    t_{14} & = & W_4 , \nonumber
\eea
when $W_1-W_2-W_3-W_4 \leq 0$. Otherwise one needs to include
bounces in which the out-edge coincides with the in-edge. In this
case a solution can be chosen as follows: \bea
    t_{11} & = & W_1 - W_2 -W_3 -W_4 , \nonumber \\
    t_{1j} & = & t_{j1} = W_j, \quad j=2,3,4,  \\
    t_{ij} & = & 0, \quad {\rm otherwise}. \nonumber
\label{rules2}
\eea

The directed-loop algorithm satisfies ergodicity as any
configuration can be obtained from another configuration by
flipping spins along a finite number of (possibly overlapping)
loops. This process is exactly the directed-loop update, thus
ergodicity follows.

The algorithm presented here has many similar features to the loop
algorithm~\cite{Evertz}. The loop algorithm breakup rules for the
symmetric eight-vertex model can be chosen identical to
Eq.~(\ref{rules}), as can be seen from Ref.~\cite{Kawashima},
whenever the weights are such that no bounces are needed in the
directed-loop algorithm. However in parameter regimes where bounces are
needed, the related feature in the loop algorithm is to ``freeze''
independent loops together. Bounces and ``freezing'' of loops are
very different in how they act to change the
configuration. While bounces is a local resistance to changing a
vertex, ``freezing'' causes big non-local changes of the vertex
configuration. There are also other differences: For general
vertex models the set of non-freezing/bouncefree solutions is
always smaller for the loop algorithm than for the directed-loop
algorithm. This allows the directed-loop algorithm to be efficient
in a larger region of parameter space than the loop algorithm. In
particular this applies to the asymmetric eight-vertex model.

Note that the need for bounces is generally not so crucial for
higher-vertex models with many weights of the same magnitude, thus
we expect that the directed-loop algorithm should work well in
simulating these. Note also that an algorithm based on the
directed-loop idea was recently demonstrated to be effective in
simulating classical integer-valued link-current
models~\cite{Alet}.

\section{Six-vertex model with the DWBC}
The six-vertex model with the DWBC was introduced in
Ref.~\cite{K-82} in connection with the calculation of the
correlation functions for exactly solvable $1+1$ dimensional
models~\cite{KBI-93}.
\begin{figure}
\begin{center}
\includegraphics[clip,width=7cm]{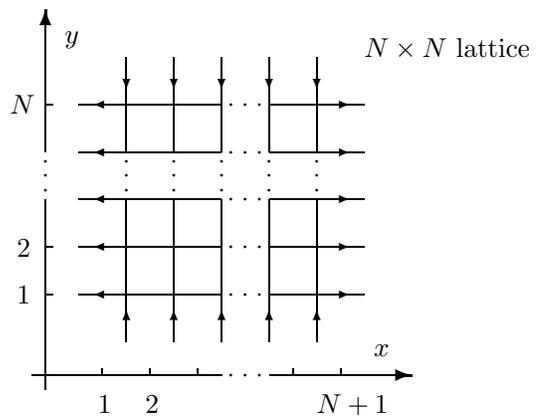}
\end{center}
\caption{The domain wall boundary conditions. Shown is an $N
\times N$ lattice. The total number of vertices is $N^2$. The $x$
and $y$ coordinates take integer values at the midpoints of the
horizontal edges. \label{DWBC}}
\end{figure}
Here we recall the definition of the model in brief, referring for
further details to the Ref.~\cite{BPZ-02} where a more detailed
description of the model and a comprehensive list of references
are given.

The model is defined on an $N\times N$ square lattice; the
thermodynamic limit corresponds to $N\to\infty.$ There are six
possible states at each vertex: one should set $d=0$ in the
eight-vertex model defined above, Fig.~\ref{vertices}. The model
is symmetric: the statistical weights, $a,$ $b,$ and $c,$ of the
allowed states are assumed to be invariant under the simultaneous
reversal of all arrows. Hence, the model is characterized by only
two parameters, which can be taken to be $a/c$ and $b/c$. We set
$c=1$ henceforth.

The DWBC imply that all arrows on the top and bottom of the
lattice are pointing inward, while all arrows on the left and
right boundaries are pointing outward, Fig.~\ref{DWBC}.

To investigate the spatially inhomogeneous behavior of this model
we focus on the polarization, $\chi_N(x,y)$~\cite{BPZ-02,BKZ-02},
which is the ensemble average of the arrow direction on the edge
with coordinates $(x,y)$ on the $N \times N$ lattice. The
coordinate system used is shown in Fig.~\ref{DWBC}. Due to the
symmetry of the model it is sufficient to consider the
polarization of the horizontal arrows only. The value $+1$ ($-1$)
is assigned to an arrow pointing to the right (left) and the
ensemble average is assumed to be normalized by dividing by the
partition function. Therefore, $\chi_N$ lies between $-1$ and $1.$

\begin{figure}
\begin{center}
\includegraphics[clip,width=5cm]{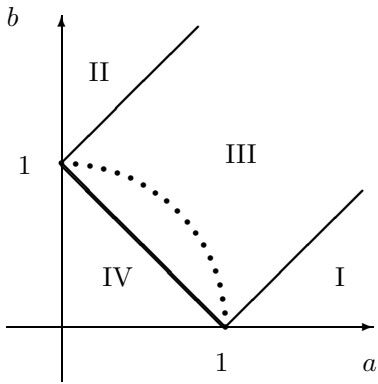}
\end{center}
\caption{The phase diagram of the six-vertex model in terms of the
weights $a$ and $b$. One has $\Delta>1$ in the regions
$\mathrm{I}$ and $\mathrm{II},$ $-1<\Delta<1$ in the region
$\mathrm{III},$ and $\Delta<-1$ in the region $\mathrm{IV}.$ The
dotted quartercircle corresponds to $\Delta=0.$}
\label{Fig:diagram}
\end{figure}

Obviously, $\chi_N$ is independent of the coordinates of the edge
in case of PBC. For these boundary conditions $\chi_N$ is known in
the thermodynamic limit, and exhibits ferroelectric order,
antiferroelectric (AF) order or no order, depending on the
position on the $(a,b)$ plane. Thus, three phases exist in the
six-vertex model with PBC: ferroelectric, antiferroelectric, and
disordered phase. In Fig.~\ref{Fig:diagram} the phase diagram on
the $(a,b)$ plane for the model with PBC is plotted (cf., Fig.~8.5
of Ref.~\cite{Baxter-82}).
 \unitlength=1pt

Introduce a parameter $\Delta$ by the formula
\begin{equation}
\Delta=\frac{a^2+b^2-1}{2ab}.
\end{equation}
The case $\Delta>1$ (regions $\mathrm{I}$ and $\mathrm{II}$ in
Fig.~\ref{Fig:diagram}) corresponds to the ferroelectric phase;
the case $-1<\Delta<1$ (region $\mathrm{III}$ in
Fig.~\ref{Fig:diagram}) to the disordered phase; the case
$\Delta<-1$ (region $\mathrm{IV}$ in Fig.~\ref{Fig:diagram}) to
the AF phase.

Fig.~\ref{Fig:diagram} may be considered as the phase diagram for
the model with the DWBC, in the sense that the free energy takes a
different analytic form in the regions $\mathrm I$ through
$\mathrm{IV}$ (see Ref.~\cite{ZJ-00} for details). But, in case of
the DWBC the polarization $\chi_N$ depends on the position of the
edge. In the next section we show numerical results for the
polarization $\chi_N(x,y)$ of the horizontal arrows as the
parameters $a$ and $b$ are varied.

Making use of the directed-loop algorithm described in the
previous section for simulation of the model with the DWBC one
should treat vertices belonging to the boundary and the bulk
vertices differently. In the bulk one finds by setting $d=W_4=0$
and $c=1$ in Eqs.~(\ref{rules}), that bounces are only necessary
when $a+b < 1$ or $|a-b| > 1$. For the boundary vertices the loop
is not allowed to exit on the boundary edges, because the arrows
on these edges are fixed by the boundary conditions. This leads to
more restricted equation sets (many $W$'s  are equal to zero) for
the boundary vertices and generally requires the inclusion of
bounce processes.

Another important point should be mentioned is that the DWBC do
not violate the ergodicity of the algorithm even though loops
which wind around the boundaries are excluded. These winding loops
are needed in order to change the net polarization in the $x$- or
$y$-direction. However, one can verify that the boundary
conditions restricts the net polarization in both these direction
to be zero for any configuration, so winding loops are not
necessary to sample the full configuration space allowed by the
boundary conditions.

\section{Results}

In this section we present the results of the simulations for the
polarization $\chi_N(x,y)$ in the disordered, antiferroelectric
and ferroelectric phases.

\begin{figure}
\includegraphics[clip,width=8cm]{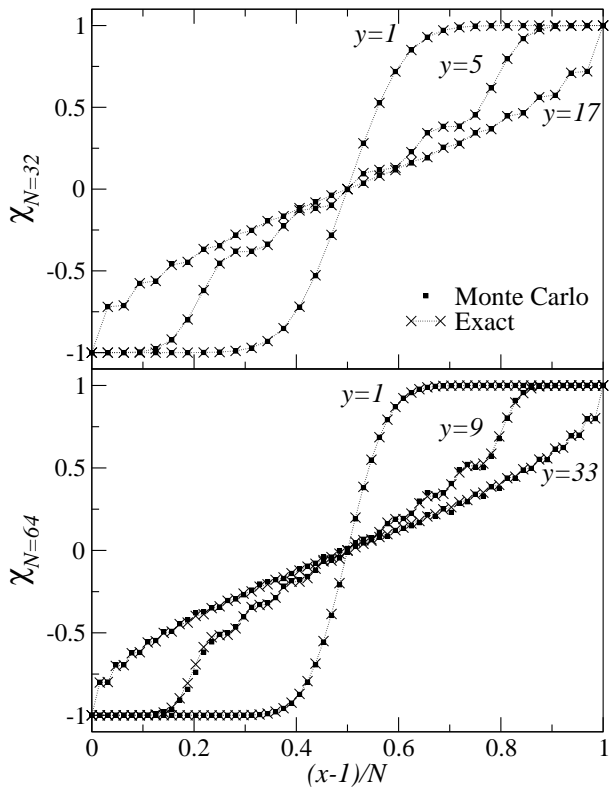}
\caption{Polarization $\chi_N(x,y)$ as a function of $x$ for
different values of $y$. Vertex weights $a=b=1/\sqrt2.$ Results
for two different system sizes are shown: $N=32$ (upper panel) and
$N=64$ (lower panel). The filled symbols are Monte Carlo results,
while the crosses are exact results gotten from Ref.~\cite{KP-04}.
The dotted lines are guides to the eye.} \label{check}
\end{figure}

(i) Disordered phase: $-1<\Delta<1.$ First consider the particular
case $\Delta=0$ (dotted quartercircle in Fig.~\ref{Fig:diagram}).
An exact expression for $\chi_N(x,y)$ in this case was obtained by
Kapitonov and Pronko~\cite{KP-04} recently. To check our algorithm
we have compared results for the polarization at the point
$a=b=1/\sqrt{2}$ with the exact results of Ref.~\cite{KP-04}. The
comparison can be seen in Fig.~\ref{check},
where the polarization is shown as a function of $x$ for different
values of $y$ and system sizes, $N$. One can clearly see that the
boundary values of the polarization, $\pm 1,$ extends a finite
distance into the bulk and this distance depends on $y.$ The areas
where the polarization stays at its boundary values are termed
``frozen'' regions. Going further into the bulk, there is a
transition to a ``disordered'' region, where apart from small
wiggles due to the finite system size, the polarization changes
smoothly. It is interesting to note that there never is any
extended regime where the polarization is zero, as is the case for
PBC. The transition between the ``frozen'' and ``disordered''
regions gets sharper as the system size is increased, as can be
seen by comparing the two panels in Fig.~\ref{check}.

\begin{figure}
\begin{center}
\mbox{
\includegraphics[clip,width=4cm]{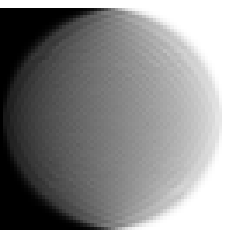}
\includegraphics[clip,width=4cm]{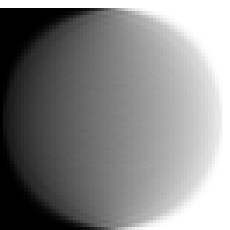}
} \hskip2cm (a) \hskip4cm (b)
\mbox{
\includegraphics[clip,width=4cm]{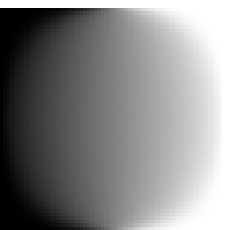}
\includegraphics[clip,width=4cm]{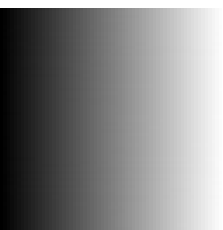}
} \hskip2cm (c) \hskip4cm (d)
\includegraphics[clip,width=6cm]{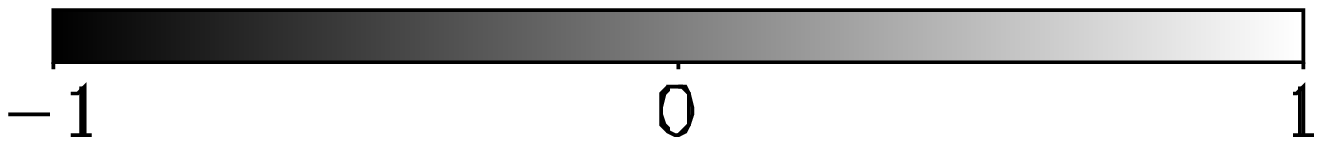}
\caption{Greyscale plot of the polarization $\chi_N(x,y)$ for
$N=64$ in the disordered phase. Vertex weights are equal, $a=b,$
and run through the values $1/\sqrt{2},$ $1,$ $3,$ $100$ for
figures (a)--(d), respectively. The corresponding values of
$\Delta$ are $0,$ $1/2,$ $17/18,$ $1-5\cdot
10^{-5}.$\label{F:disordered}}
\end{center}
\end{figure}

It is convenient to visualize the behavior of the polarization
using greyscale plots, where greyvalues are assigned to values of
$\chi_N(x,y)$ and each point $(x,y)$ corresponds to a location of
the midpoint of a horizontal edge following the layout described
in Fig.~\ref{DWBC}. For $a=b=1/\sqrt{2}$ such a plot is shown in
Fig.~\ref{F:disordered}(a).
The four ``frozen'' corners are clearly apparent. In these
regions, the vertices are all of the same type, and are, from
upper left to bottom right, $a_1$, $b_1$, $b_2$, $a_2$,
respectively. To measure the area of the ``frozen'' regions, we
define a threshold value $\epsilon = 0.08$, such that points
$(x,y)$ where $|\chi_N(x,y)| > 1-\epsilon$ are judged to be in a
``frozen'' region. Applying this we find that each ``frozen''
corner is 4.6\% of the total area. This value changes relatively
little changing the value of $\epsilon$.

Going away from the $\Delta=0$ curve, let us follow along the
diagonal, $a=b,$ towards $\Delta=\infty$ first,
Fig.~\ref{F:disordered}. As the values of the vertex weights $a$
and $b$ increase, the area of the ``frozen'' regions decreases. We
find that with $\epsilon=0.08$ each frozen corner in $(b)$ is
$4.0\%$ of the total area, and $2.8\%$ in (c). For very large
values of $a=b$, the polarization $\chi_N(x,y)$ increases linearly
from $-1$ to $1$ as $(x-1)/N$ goes from $0$ to $1$, independent of
$y$, as can be seen in Fig.~\ref{F:disordered} (d). This is
consistent with what is expected from an ensemble of
configurations with the smallest possible number of $c$-type
vertices: $N!$ configurations each with a single $c$-type vertex
on every row and column.
\begin{figure}
\begin{center}
\mbox{
\includegraphics[clip,width=4cm]{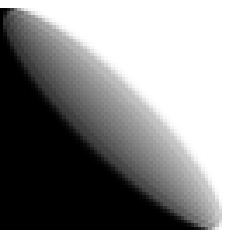}
\includegraphics[clip,width=4cm]{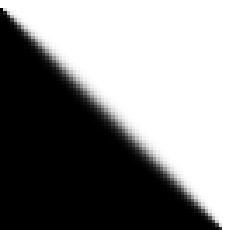}
} \hskip2cm (a) \hskip4cm (b)

\includegraphics[clip,width=6cm]{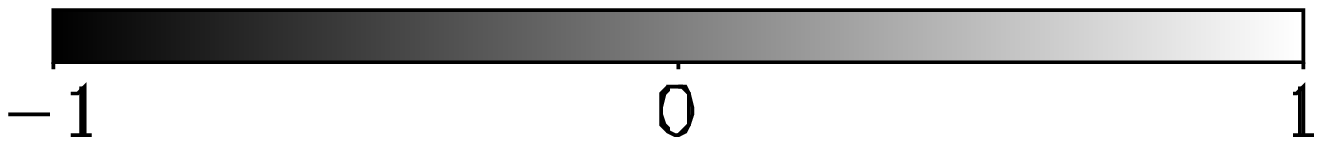}
\end{center}
\caption{Greyscale plot of the polarization $\chi_N(x,y)$ for
$N=64.$ The weight $a=1/4,$ while the weight $b$ is chosen to be
$b=\sqrt{15/16}$ ($\Delta=0,$ disordered phase) in figure (a) and
$b=5/4$ ($\Delta=1,$ the boundary between disordered and
ferroelectric phases) in figure (b). \label{Fig:disordered2} }
\end{figure}

Consider now $a \neq b.$ Because of the symmetry of the phase
diagram, Fig.~\ref{Fig:diagram}, one can choose $b>a$ without loss
of generality. The weights of the vertices in the four ``frozen''
corners are no longer equal, and the ``disordered'' region
distorts into an oblong shape oriented along the diagonal with
large corners of $b_2$ and $b_1$ vertices, see
Fig.~\ref{Fig:disordered2}. The simulations for $a=1/4$ and
$b=\sqrt{15/16}$ are shown in Fig.~\ref{Fig:disordered2}(a). The
width of the oblong region shrinks as $b$ increases keeping $a$
fixed, $a=1/4$, and becomes very thin at the boundary to the
ferroelectric region, as can be seen in
Fig.~\ref{Fig:disordered2}(b). Along this boundary, $b=a+1$, the
width of the oblong region expands as $a$ increases with $N$ being
constant.

\begin{figure}
\begin{center}
\mbox{
\includegraphics[clip,width=4cm]{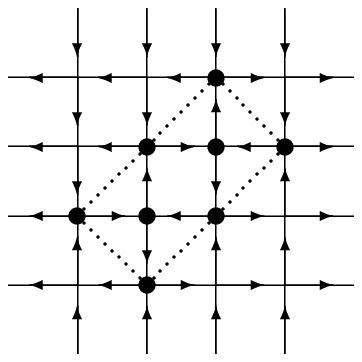}
\includegraphics[clip,width=4cm]{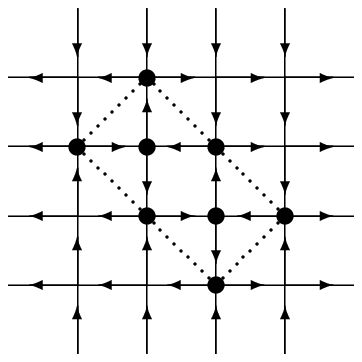}
}
\end{center}
\caption{The two configurations having maximum number of the
$c$-type vertices. These vertices are marked by filled circles.
The size of the lattice is $4\times 4.$ \label{Fig:af}}
\end{figure}

\begin{figure}
\begin{center}
\mbox{
\includegraphics[clip,width=4cm]{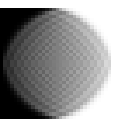}
\includegraphics[clip,width=4cm]{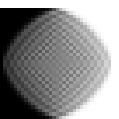}
} \hskip2cm (a) \hskip4cm (b)

\mbox{
\includegraphics[clip,width=4cm]{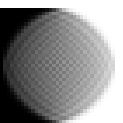}
\includegraphics[clip,width=4cm]{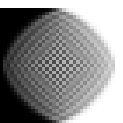}
} \hskip2cm (c) \hskip4cm (d)

\includegraphics[clip,width=6cm]{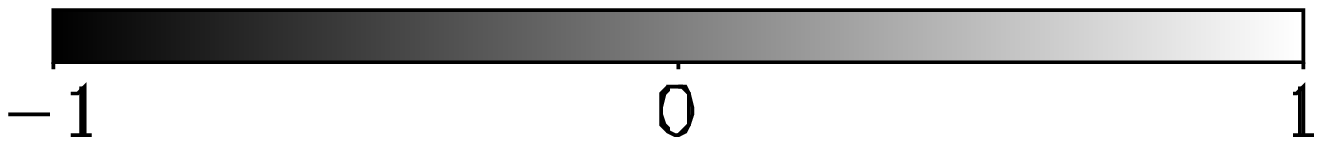}
\caption{Greyscale plot of the polarization $\chi_N(x,y)$ for two
different system sizes: $N=32$ in figures (a) and (b) and $N=33$
in figures (c) and (d). Vertex weights are equal, $a=b,$ and take
the value $1/2$ ($\Delta=-1,$ the boundary between disordered and
AF phases) for figures (a) and (c), and the value $3/8$
($\Delta=-23/9,$ AF phase) for figures (b) and (d).
\label{Fig:afcase} }
\end{center}
\end{figure}

(ii) Antiferroelectric phase: $\Delta<-1.$ The simulations in the
AF phase are less efficient than in the disordered phase. This is
partly due to the presence of the bounce processes also for bulk
vertices, but another feature which makes the simulations
difficult in this phase is the degeneracy of the two types of AF
orders. In the AF phase it becomes energetically favorable to have
a maximum possible amount of $c$-type vertices, which is achieved
by placing $c$-type vertices in a diamond placed in the center of
the lattice. For an even $N$ this diamond can be placed in two
equivalent places differing only by one lattice spacing, as shown
in Fig.~\ref{Fig:af}.
The Monte Carlo algorithm is however slow in tunnelling between
these configurations, and this sets a limit to its performance.
For odd $N$ there is no such a degeneracy and the simulations are
more efficient. Greyscale plots of the polarization for $a=b=1/2$
and $a=b=3/8$ are shown in Fig.~\ref{Fig:afcase}.
We have plotted results for both even and odd $N.$

One can see that the ``disordered'' region have a diamond-like
shape, which is consistent with the domination of the $c$-type
vertices in the AF phase. As $a=b$ decreases $(\Delta \to
-\infty)$, the shape of the ``disordered'' region should converge
to the one shown in Fig.~\ref{Fig:af}, that is, the boundaries of
the ``disordered'' region should become more and more straight.
But, this convergence appears to be rather slow and it
is not easy to see it from Fig.~\ref{Fig:afcase}. What one can
clearly see from Fig.~\ref{Fig:afcase} is the difference between
odd and even $N.$ For odd $N$ AF oscillations are clearly visible
in the center of Figs.~\ref{Fig:afcase}(c) and (d), while they are
much weaker for even $N$, Figs.~\ref{Fig:afcase}(a) and (b),
reflecting the degeneracy mentioned above. These difference
between even and odd $N$ can also be clearly seen from
Fig.~\ref{Fig:evenoddosc}. For odd $N$ AF oscillations are weaker
at $a=b=1/2$ than at $a=b=3/8$.

\begin{figure}
\begin{center}
\mbox{
\includegraphics[clip,width=4cm]{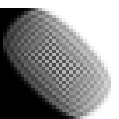}
\includegraphics[clip,width=4cm]{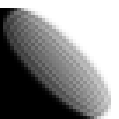}
} \hskip2cm (a) \hskip4cm (b)
\includegraphics[clip,width=6cm]{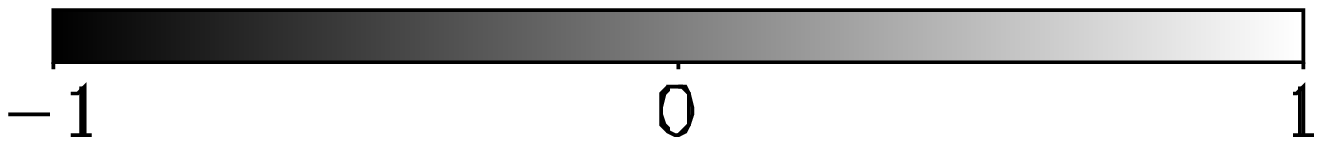}
\end{center}
\caption{Greyscale plot of the polarization $\chi_N(x,y)$ for
$N=32.$ The weight $a=1/4$ while the weight $b$ is chosen to be
$b=1/2$ ($\Delta=-11/4,$ AF phase) in figure (a) and $b=3/4$
($\Delta=-1,$ the boundary between disordered and AF phases) in
figure (b). \label{something}}
\end{figure}

For $a \neq b$ greyscale plots are shown in Fig.~\ref{something}.
Here AF oscillations in the middle of the plot are visible for
$a=1/4$ and $b=1/2$, Fig.~\ref{something}(a), while they have
almost vanished at the boundary between the AF and disordered
phases, Fig.~\ref{something}(b).

(iii) Ferroelectric phase: $\Delta>1.$ The behavior of the
polarization in this phase is essentially the same as shown in
Fig.~\ref{Fig:disordered2}. Vertices of type $b$ dominate
completely in the region $\mathrm{II}$ of the phase plane
Fig.~\ref{Fig:diagram}, while in the region $\mathrm I$ of the
phase plane the dominant vertices are those of type $a.$ If one
goes along the phase boundary, $b=a+1,$ towards $a=\infty,$ the
widths of the ``disordered'' region is increased, as we have
mentioned in the end of the part (i) of this Section.

The exact expression is known~\cite{BPZ-02} for the polarization
along the boundary, $\chi_N(x,1).$ Comparing our Monte Carlo data
to this expression we find that in no cases is the absolute
difference bigger than $0.016,$ which is comparable to the
statistical errors of our simulations.

\section{Discussion}

\begin{figure}
\begin{center}
\includegraphics[clip,width=8cm]{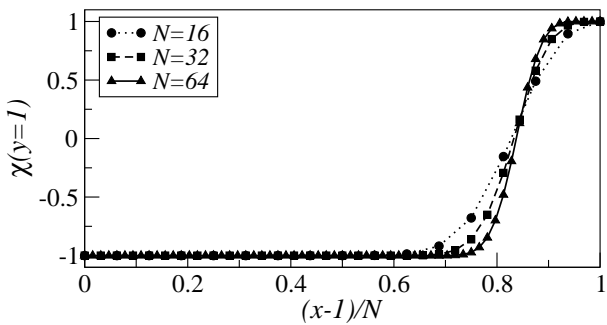}
\end{center}
\caption{Boundary polarization $\chi_N(x,1)$ is shown for three
system sizes, $N=16$, $32$, and $64$. Vertex weights $a=1/4$ and
$b=3/4$ ($\Delta=-1,$ the boundary between disordered and AF
phases). Note the steepening of the curve as $N$ increases.}
\label{Fig:boundarypol}
\end{figure}

We have considered the phase diagram of the model for the given
$N.$ Now, discuss the following problem: what happens with
$\chi_N(x,y)$ in the thermodynamic limit, $N\to\infty?$ It is
natural to expect that differences in the behavior of the
polarization in the different parts of the phase plane,
Fig~\ref{Fig:diagram}, become more pronounced as $N\to\infty.$ As
one can see in Fig.~\ref{check}, the wiggles in the ``disordered''
region decrease with $N$ increasing, and this is, indeed, the case
for all the points ($a,b$) lying in the disordered phase
($-1<\Delta<1,$ region $\mathrm{III}$ of the phase plane,
Fig.~\ref{Fig:diagram}) and checked in our simulations. We expect
that these wiggles, coming from the antiferroelectrically ordered
configurations, should vanish completely in the thermodynamic
limit for this phase. The next conjecture we want to make is on
the behavior of the polarization along the boundary,
$\chi_N(x,1).$ It is known that for $\Delta=0,$ as well as at the
point $a=b=1,$ the boundary polarization becomes the Heaviside
step function in the thermodynamic limit~\cite{BPZ-02,Z-96}. We
conjecture that this is the case for the whole disordered phase;
the position of the discontinuity will depend on the ratio between
$a$ and $b.$ We present Fig.~\ref{Fig:boundarypol} to support
this conjecture.

Furthermore, note that for $a=b=1/\sqrt2$ there is a mapping (see,
{\it e.g.}, Ref.~\cite{ZJ-00}) of the six-vertex model with the
DWBC onto the so-called model of domino tilings of the Aztec
diamond. The thermodynamic behavior of the latter model was
investigated in Refs.~\cite{JPS-98}. It shows the same features as
in Fig.~\ref{F:disordered}(a): the tilings are ordered (frozen) in
the corners of the diamond, while going inside one falls into the
``disordered'' region. All these features were named the ``Arctic
Circle Theorem'', since the shape of the boundary between the
``frozen'' and ``disordered'' regions is circular. The transition
between ``frozen'' and ``disordered'' regions is step-like, with
the height of the step function depending on the coordinates $x$
and $y.$

We expect the analogue of the Arctic Circle Theorem to take place
for the whole disordered phase, $-1<\Delta<1$: there should be the
``frozen'' regions, ``disordered'' region, and a sharp transition
between them. We expect also that the profile of the boundary
between the ``frozen'' and ``disordered'' regions is circular for
$a=b$, even though there is no obvious symmetry protecting this
statement. Note that the very ``smeared'' profile in
Fig.~\ref{F:disordered}(d) does not contradict our hypotheses
because $N=64$ is relatively small compared to the values of the
vertex weights $a$ and $b,$ and is thus far from the thermodynamic
limit for this point of the phase diagram.

\begin{figure}
\begin{center}
\includegraphics[clip,width=8cm]{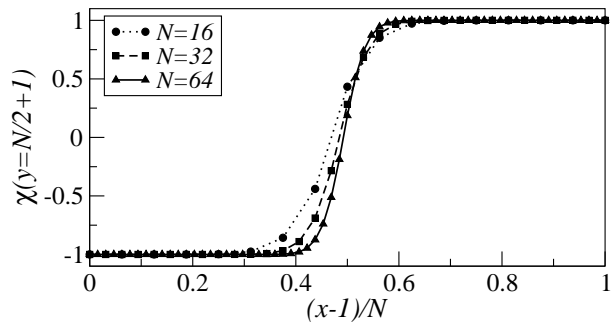}
\end{center}
\caption{Polarization $\chi_N(x,y=N/2+1)$ is shown for three
system sizes, $N=16,$ $32,$ and $64.$ Vertex weights $a=1/4$ and
$b=5/4$ ($\Delta=1,$ the boundary between disordered and
ferroelectric phases). Note the steepening of the curve as $N$
increases. \label{Fig:ferroprofile}}
\end{figure}

For the ferroelectric phase, $\Delta>1,$ the greyscale plot
Fig.~\ref{Fig:disordered2}(b) together with the scans shown in
Fig.~\ref{Fig:ferroprofile} leads to the natural conjecture: in
the whole region $\mathrm{II}$ of the phase plane,
Fig.~\ref{Fig:diagram}, a sharp discontinuity from a ``frozen''
domain with $b_1$ vertices to the one with $b_2$-vertices takes
place in the thermodynamical limit. In the region $\mathrm{I}$ the
behavior is essentially the same, one should simply use $a$-type
vertices instead of the $b$-type.

\begin{figure}
\begin{center}
\includegraphics[clip,width=8cm]{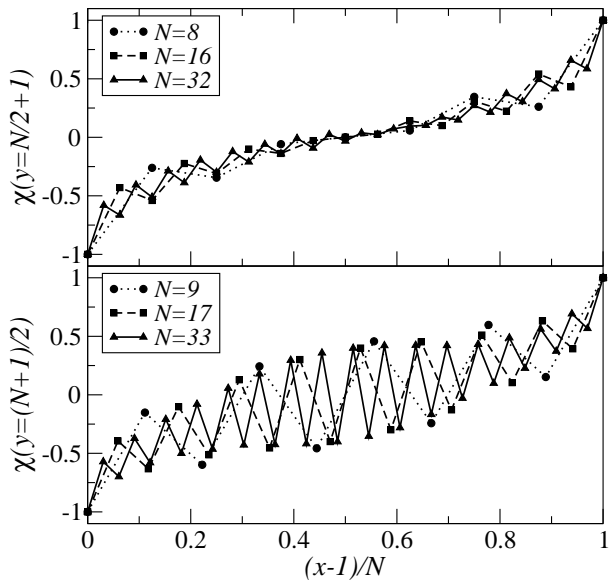}
\end{center}
\caption{Polarization $\chi_N(x,y)$ along lines of constant $y$,
where $y=N/2+1$, $y=(N+1)/2$) for even and odd $N$ respectively,
 is shown for $N=8,$ $16,$
and $32$ (upper panel) and $N=9,$ $17,$ and $33$ (lower panel).
Vertex weights $a=b=3/8$ ($\Delta=-29/9,$ AF phase).
\label{Fig:evenoddosc}}
\end{figure}

To this end, consider the antiferroelectric phase, $\Delta<-1$. We
expect the step-like behavior of the boundary polarization,
$\chi_N(x,1)$ in this phase in the thermodynamic limit, as well as
the existence of the ``frozen'' regions in the corners. Our
statements on the behavior of the polarization deep inside the
lattice are more speculative. For $a=b$ and even $N$ the height of
the AF oscillations decreases, while for odd $N$ these
oscillations seem to be non-vanishing in the large $N$ limit, see
Fig.~\ref{Fig:evenoddosc}.
Our belief is that there is a finite region with AF order for odd
$N,$ as $N\to\infty,$ while for even $N$ the polarization exhibits
no such an order.

Finally, we would like to stress that the directed-loop algorithm
can also be applied to study the six-vertex model with any
boundary conditions, and the higher-vertex models. These could
help in solving the problems for which the analytical methods are
difficult to apply. For example, the six-vertex model with any
boundary conditions can be considered as a model for a description
of interface roughening of a crystal surface \cite{Beij-77}. An
important point in studies Refs.~\cite{Beij-77} is the existence
of exact analytical results for the six-vertex model with
PBC~\cite{Lieb,Baxter-82}. Therefore, numerical data referring
to other boundary conditions than PBC could give
a new insight for these studies.

\begin{acknowledgments}
We thank V.V.~Cheianov and A.G.~Pronko for useful discussions and
the authors of the work \cite{KP-04} for providing us with their data.
M.B. Zvonarev's work was supported by the Danish Technical
Research Council via the Framework Programme on Superconductivity.
Monte Carlo calculations were in part carried out using NorduGrid,
a Nordic facility for Wide Area Computing and Data Handling.
\end{acknowledgments}

\end{document}